\documentclass[aps,prd,twocolumn,showpacs,preprintnumbers,amsmath,amssymb]{revtex4}
\usepackage{graphicx}
\usepackage{dcolumn}
\usepackage{bm}

\begin{document}

\title{Sums over geometries and improvements on the mean field approximation}

\author{Vincent E. Sacksteder IV}
 \email{vincent@authors-last-name.com}
 \homepage[]{www.sacksteder.com}
 \affiliation{S.N. Bose National Centre for Basic Sciences, Sector III, Block JD, Salt Lake, Kolkata 700098, India.  }

\date{\today}

\begin{abstract}
The saddle points of a Lagrangian due to Efetov are analyzed.  This Lagrangian was originally proposed as a tool for calculating systematic corrections to the Bethe approximation, a mean-field approximation which is important in statistical mechanics, glasses, coding theory, and combinatorial optimization.  Detailed analysis shows that the trivial saddle point generates a sum over geometries reminiscent of dynamically triangulated quantum gravity, which suggests new possibilities to design sums over geometries for the specific purpose of obtaining improved mean field approximations to $D$-dimensional theories.  In the case of the Efetov theory, the dominant geometries are locally tree-like, and the sum over geometries diverges  in a way that is similar to quantum gravity's divergence when all topologies are included. Expertise from the field of dynamically triangulated quantum gravity about sums over geometries may be able to remedy these defects and fulfill the Efetov theory's original promise.  The other saddle points of the Efetov Lagrangian are also analyzed; the Hessian at these points is nonnormal and pseudo-Hermitian, which is unusual for bosonic theories.  The standard formula for Gaussian integrals is generalized to nonnormal kernels.
\end{abstract}


\maketitle

\section{INTRODUCTION}
Modern physics often struggles with geometry.  Zero dimensional and "infinite dimensional" systems are the simplest to analyze, precisely because in such systems the effects of dimension and distance are grossly simplified.  In this paper we will restrict our attention to the more realistic and challenging scenario: a finite number of dimensions $D$.  In this scenario physicists typically choose ground states - saddle points - that are spatially uniform, which means that we are pretending that the system "almost" lacks spatial structure.  Geometrical information is later added back into perturbative calculations via the bare Green's function.  This program of mean field theory fails when the number of dimensions is small, and in disordered and complex systems often must be augmented with replicas and replica symmetry breaking \cite{Parisi87}. It would be desirable to find a systematic perturbative expansion which incorporates geometrical effects non-perturbatively; in the zeroth-order approximation.  Such a theory would include incomplete geometrical information in the lowest order, and would systematically improve its geometrical accuracy at higher orders.  

One very tempting possibility is to design an expansion whose zeroth order is the Bethe approximation \footnote{The Bethe approximation is closely related to the TAP equations and the cavity approximation \cite{Parisi87}.  It also is the foundation of the belief propagation and message passing algorithms which have proved useful in coding theory and in combinatorial optimization problems \cite{Kabashima98, Yedidia01}. }. Bethe required that if a spin residing at node $a$ is removed then the spins on neighbors of $a$ will be uncorrelated \cite{Bethe35, Peierls36}.   On a tree his approximation is exact; thus its physical meaning consists of including complete information about nearest neighbors, but leaving out the effects of loops, i.e. possibilities to make round trips without retracing one's steps.  When treating low-dimensional, disordered, or frustrated systems, the Bethe approximation gives notable improvements over mean field theory \cite{Parisi87}.  It is equivalent to using the replica method without breaking replica symmetry, and has a hierarchy of simple generalizations which are equivalent to successive levels of replica symmetry breaking \cite{Parisi87}. 

A systematic expansion of corrections to the Bethe approximation would adjust for the difference between the correct D-dimensional geometry and a tree.  Several authors have proposed improvement schemes based on higher correlated moments \cite{Montanari05}, effects of nearby spins on one another \cite{Kikuchi51, Georges91}, or  "generalized loops" moving around the correct geometry \cite{Chertkov06a, Chertkov06b}.  All of these schemes involve an exponentially large number of corrections, as is typical in perturbative calculations.  Less typically, in no formalism has anyone identified a small parameter whose powers can be used to justify truncating the corrections.  The crucial problem for the Bethe approximation today is that of finding a small parameter which controls its corrections.

In 1990 Efetov \cite{Efetov90} proposed a new formalism for calculating corrections to the Bethe approximation. Much more recently Parisi and Slanina \cite{Parisi06} worked out certain details of applying the new scheme to a D-dimensional geometry.   The advantage of Efetov's formalism is that it uses the standard mathematical machinery of saddle point approximations, while the geometrical details of the Bethe approximation and its corrections have somehow been hidden in a clever choice of Lagrangian.   These authors claimed that the new Efetov Lagrangian: 
\begin{itemize}
 \item reproduces the Bethe approximation as the zeroth order of a saddle point approximation.
\item 
  is able to reproduce the partition function of any statistical model with two-body interactions, via suitable choices of the Efetov Lagrangian's parameters.  For example, the Efetov Lagrangian can reproduce any Potts model, including the Ising model, as well as any lattice gauge theory.  It is important to keep firmly in mind the distinction between the Efetov Lagrangian itself and the "target theory" - the model which the Efetov Lagrangian is being tuned to reproduce, accompanied by a complete specification of the geometry.
\end{itemize}
The second assertion, giving the Efetov theory's physical meaning, was based on a saddle point expansion around the trivial saddle point. The first assertion, giving the connection with the Bethe approximation, was based on a saddle point expansion around a different, non-trivial, saddle point \footnote{There was an implicit assumption, which will not be discussed further here, that the two saddle point expansions must converge to the same value.}.

In this paper, section \ref{SumOverGeometries} questions Efetov, Parisi, and Slanina's second assertion, that the Efetov Lagrangian can replicate the partition function of any target theory.  It reports that while they are correct that one of the corrections to the trivial saddle point is the target theory's partition function, this correction is only one of many.  The other (newly discovered) corrections dominate over the target partition function, contain two or more spins for each spin in the original target theory, and correspond to a sum over geometries reminiscent of dynamically triangulated quantum gravity \cite{Ambjorn95, Loll98, Ambjorn06}, but without the restriction to simplexes having triangular faces.  Even though the target theory which the Efetov Lagrangian is trying to replicate lives on a cubic lattice with D dimensions and N nodes, the dominant corrections to the trivial saddle point live on random geometries with many nodes and no specific dimension.  The dominant random geometries are locally tree-like, which perhaps explains why the Bethe approximation emerges from the non-trivial saddle points.   

This is a very illuminating result.  It reveals that improving on mean field theory in $D$ dimensions is a matter of defining an appropriate sum over geometries (preferably using an integral formulation), and analyzing it with techniques from statistical mechanics and quantum gravity.  If the relative importance of $D$-dimensional lattice geometries vs. others can be tuned smoothly, then a mean field theory with systematic corrections can be developed \footnote{For theories with a Lagrangian structure similar to the Efetov theory, it may be possible to force lattice dominance by adding a term to the action proportional to the number of short self-avoiding loops.}.  Assuming that smooth reweighting can arrive at a regime where tree-like geometries dominate, the mean field theory would likely be governed by the Bethe approximation.  Even if smooth reweighting is not possible, any non-trivial saddle points will offer an approximate mean field theory with systematic corrections as long as lattice geometries are dominant.  

As it stands Efetov's theory does not fulfill its promise: lattice geometries do not dominate.  Moreover, the sum over geometries grows factorially with the volume, not exponentially, and is therefore in need of regularization. This is a familiar problem from quantum gravity and string theory \cite{Ambjorn95, Loll98}, where almost universally the sums over geometries with different topologies diverge, and is a sign of a nonextensive entropy \cite{Tsallis02}.  See \cite{Canfora06} for a recent discussion of the link between nonextensive entropy, string theory, and replica symmetry breaking in frustrated systems.  

Sometimes a failed (divergent) theory is a success, if it points the way to other better theories, or gives physical insight into a problem - this is the motivation for continued research in quantum gravity and strings.  The Efetov theory is a similar sort of success: it suggests that a similar sum over geometries, suitably regularized using expertise from quantum gravity and string theory, could provide a systematic scheme for improving on the Bethe approximation.  

The key technical question is whether a suitable regularization can be found.   Efetov's theory considers only discrete geometries; therefore a lattice cutoff is useless.  Alternative regularizations like those found in dynamically triangulated quantum gravity may be useful.  Topological regularizations \cite{Ambjorn95, Loll98} have found some success especially in $1+1$ dimensions - these restrict the sum over geometries to planar geometries (topological spheres with no holes), or tori (a single hole), etc.  More recently, another restriction on the sum over geometries - Causal Dynamical Triangulations - has been shown to successfully regularize quantum gravity \cite{Ambjorn06}.   There is good reason to hope that physicists in quantum gravity and string theory, armed with their expertise in designing and analyzing sums over geometries, could find a sum over geometries suitable for calculating corrections to the Bethe approximation.  They now have the possibility of making a direct and immediate impact on low energy physics, i.e. condensed matter theory.

In other words, the discovery that the Efetov theory is a sum over geometries suggests that the Bethe approximation - the most powerful mean field theory in existence - is best understood from a perspective where geometry is not pre-determined or static, but a physical variable.  From this perspective, the Bethe approximation is the low energy limit of some - as yet unknown - sum over geometries similar to those found in quantum gravity.  Perhaps the historical development of the Bethe approximation via heuristic arguments was both fortuitous and accidental; perhaps future physicists will find that rigorous derivation of the Bethe approximation is an exercise in quantum gravity.  In my opinion, this new perspective is the most important contribution of the present paper.  For a parallel perspective, see Benedetti and Loll's recent argument that averaging over geometries may improve the analysis of critical behavior in D dimensions by removing lattice artifacts \cite{Benedetti06a, Benedetti06b}.

Section \ref{BetheSaddlePoint} turns to the non-trivial saddle points - the ones producing the Bethe approximation and its perturbative corrections.  Where Efetov, Parisi, and Slanina concentrated on developing the perturbation theory around the saddle points, here the focus is on a full characterization of the saddle points themselves.  Section \ref{BetheSaddlePoint} begins by showing for the first time how to integrate out the exact symmetries of the Efetov Lagrangian, a challenge which Parisi and Slanina had noted but not solved.   It  next works out the details of the saddle points and their Hessian, giving special attention to a novel mathematical challenge which was not discussed by Efetov, Parisi, and Slanina: their theory's Hessian and bare Green's function are nonnormal, meaning that ${[H, H^{\dagger}]} \neq 0$ \cite{Golub83}.  Such matrices have distinct sets of right and left eigenvectors, and do not obey many of the usual theorems from linear algebra.  In particular, the right eigenvectors are not orthogonal to each other, and the  same applies to the left eigenvectors.  Nonnormal matrices have been of particular scientific interest in describing non-equilibrium and chaotic phenomena \cite{Trefethen05} and also dissipation \cite{Bender98, Mostafazadeh02}, but I am not aware of any previous discussion of nonnormal bosonic field theory.     Lastly, section \ref{Outlook} summarizes the principal contributions of this paper and discusses possibilities for further research.

\section{\label{SumOverGeometries} THE TRIVIAL SADDLE POINT IS A SUM OVER GEOMETRIES}

Efetov, Parisi, and Slanina discuss a class of target theories $T$:
\begin{itemize}
\item The geometry is described by a graph $G$, with $V$ nodes and $L$ links.
\item Degrees of freedom reside on nodes of $G$.
\item The interactions are restricted to pairwise (two body) interactions.
\end{itemize}

Efetov, Parisi, and Slanina propose a field theory which they claim reproduces the partition function $Z^T$ of any such target theory $T$.  More specifically, they claim that $Z^T$ is equal to
\begin{equation} \label{EfetovTheory} Q \equiv {{\lambda}^{-V} {[ {\ln{\int {d\psi}{d \overline{\psi}} e^{{\mathcal{L}}_{0} + {\lambda \mathcal{L}_{1}}} }}  - {\ln{\int {d\psi}{d \overline{\psi}} e^{{\mathcal{L}}_{0} } }}   ]} }.
\end{equation}

If the target theory inhabits a $D$-dimensional cubic lattice, the Efetov Lagrangian \footnote{Efetov, Parisi, and Slanina do not include the perturbative parameter $\lambda$ in their theory, which is equivalent to setting it equal to $1$.} is
\begin{eqnarray}
\label{EfetovLagrangian}
\nonumber \mathcal{L} = {{{\mathcal{L}}_{0}} + {\lambda {\mathcal{L}}_{1}}}  =  &-& {\sum_{v d s_1 s_2} {\overline{\psi}}_{v d s_1} {\psi}_{v d s_2} {({\gamma}^{-1})}_{s_1 s_2}} 
\\ &+& {\lambda {\sum_{vs} {\xi}_{s} \;{{\Pi}_d{\psi}_{vds} {\overline{\psi}}_{v-\hat{d}, ds}}}}.
\end{eqnarray}

The complex variables $\psi$ are link variables which live at each edge on the graph.  They are completely specified by three indices: the position $v$, the direction $d$ of the link, and the spin index $s$.   These $\psi$ variables are the actual degrees of freedom in the Efetov model, and are completely distinct from the degrees of freedom of the target theory, which are encoded in the sums over the spin index $s$.  $\gamma$ is a matrix describing the interaction between two spin degrees of freedom in the target theory; for the Ising model $\gamma_{s_1 s_2} = \exp{(\beta s_1 s_2)}$.  Similarly, ${\xi}_{s}$ is a factor encoding the magnetic field.  The notation $v - \hat{d}$ used in equation \ref{EfetovLagrangian} specifies a particular neighbor of vertex $v$: namely the one which is reached by going along the $d$ axis in the negative direction. The notation adopted here assumes that the target theory lives on a cubic lattice, but suitable notations can be easily found for other target theories.

In order to understand the physical content of this Lagrangian, Efetov, Parisi, and Slanina performed a perturbative expansion around the $\psi = \overline{\psi} = 0$ saddle point \footnote{Note that there is no such saddle point if the target theory inhabits a graph which contains dangling nodes; however it is possible that such theories have a saddle point solution with values exponentially close to $0$ on nodes which are far from the dangling nodes. It is also worth noting that the trivial saddle point is not a local maximum of the action unless $\gamma$ is positive definite.  In the Ising model this amounts to requiring a ferromagnetic coupling.}.  Parisi and Slanina \cite{Parisi06} stated that the perturbative expansion around the trivial saddle point contains a Feynman diagram exactly equal to the partition function of the original theory plus diagrams equal to the "partition function of the same model wrapped several times around."  They expect that "in the thermodynamic limit the wrapping is unessential" because the wrapped geometries differ from the target geometry only at their boundaries.

The main point of this section is to show that the perturbative expansion around the trivial saddle point is actually a sum over many geometries differing from the target theory throughout their volumes, and to analyze the physics of this sum. In fact $Q$ is equal to
\begin{equation} {\sum}_{r=1}^{\infty}{\lambda}^{V{(r - 1)}} Z^T_r.\end{equation}
The first contribution to this sum, $Z^T_1$, is the desired partition function of the target theory $T$.  However, there are also infinitely many other contributions which cannot be neglected: these terms $Z^T_r$ are partition functions of theories with $r$ times as many vertices as the target theory.   For instance, if the target theory inhabits a graph with $100$ vertices,  then there will be terms with $200$, $300$, $400$, etc. vertices.  The only way to remove the higher order terms is to remove all loops from the graph inhabited by the target theory, thus restricting oneself to working on trees.  

\begin{figure}
 \includegraphics{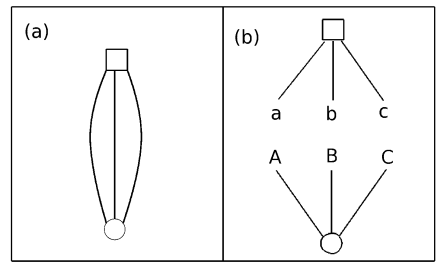}
	\caption{Part a is a simple example target geometry; node 1 is a box, node 2 a circle.  Part b: the two vertices in the perturbation theory for this target geometry.}
	\label{fig1}
\end{figure}
To show that $Q$ contains a sum over geometries, let us work out the details of a particularly simple target theory living on two nodes connected by three links, as shown in figure \ref{fig1}a.  In this case the Efetov Lagrangian is
\begin{eqnarray}
\label{EfetovLagrangianSimple}
 \mathcal{L} = {{{\mathcal{L}}_{0}} + {\lambda {\mathcal{L}}_{1}}}  =  &-& {\sum_{d=1}^3 \sum_{s_1 s_2} {\overline{\psi}}_{d s_1} {\psi}_{d s_2} {({\gamma}^{-1})}_{s_1 s_2}} 
\\ \nonumber &+& {\lambda \sum_{s}  \; {({{\xi}_{s}^1 \; {\Pi}_d{\psi}_{ds}} + {{\xi}_{s}^2 \; {\Pi}_d {\overline{\psi}}_{ds}})}}.
\end{eqnarray}
In the perturbative expansion ${\mathcal{L}}_{1}$ generates two vertices, each with three legs.  As illustrated in figure \ref{fig1}b, the legs are labeled; the first vertex has legs labeled a, b, and c while the second vertex has legs labeled A, B, and C.  The Efetov Lagrangian's bare propagator connects a with A, b with B, or c with C.  In the formulas corresponding to each Feynman diagram, one puts a factor of ${\gamma}_{s_1 s_2}$ for each instance of the propagator, a $\lambda \sum_{s} {\xi}_{s}^1$ for each instance of the first vertex, and a $\lambda \sum_{s} {\xi}_{s}^2$ for each instance of the second vertex.  

\begin{figure}
\includegraphics{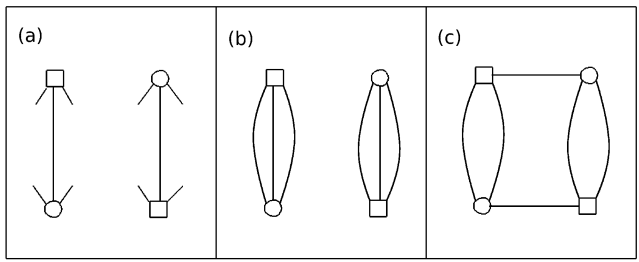}
	\caption{The second order geometries for our example.  Part a - the spanning trees. Part b - the disconnected geometry, weight $1/2$. Part c - the connected geometry, weight $3/2$.}
	\label{fig2}
\end{figure}
$Z^T_1$, the first order contribution to equation \ref{EfetovTheory}, contains one instance of each vertex, and there is only one way to connect the two, shown in figure \ref{fig1}a.  This graph is exactly the same as the target theory's geometry, so this first order correction produces exactly the partition function of the target theory.  However the second order correction $Z^T_2$ contains two instances of each vertex, and now one must think a bit.  There are ${(2!)}^3$ possible ways of connecting them; one factor of $2!$ can immediately be removed by constructing spanning trees, as shown in figure \ref{fig2}a.  We now have ${(2!)}^2$ possible choices about how to connect the four remaining pairs of legs; figures \ref{fig2}b and \ref{fig2}c show the two second order diagrams.  The first is a disconnected diagram with symmetry factor $1/2$, and is removed by the logarithm in equation \ref{EfetovTheory}.   The second diagram connects the two spanning trees, contains four nodes instead of the two nodes present in the original target theory, and has symmetry factor $3/2$.  
\begin{figure}
\includegraphics{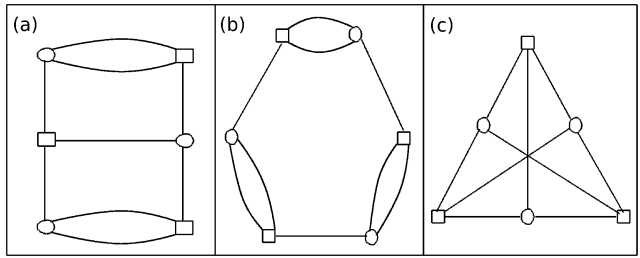}
	\caption{The third order geometries for our example.  The weights are $3$ (a), $1$ (b), and $1/3$ (c).}
	\label{fig3}
\end{figure}

The third order contribution $Z^T_3$ can be found in a similar fashion, by first constructing the spanning trees, and then enumerating the ${(3!)}^2$ possible ways of connecting the six remaining pairs of legs.  There are of course disconnected geometries that are again removed by the logarithm.  However there are also three topologically distinct fully connected geometries, which are shown in figure \ref{fig3} along with their symmetry factors.  The best way to figure out the geometries and symmetry factors reliably is to do the enumeration by hand.

Consider now evaluating $Z^T_r$ for a theory inhabiting a more complicated target graph with $V$ nodes and $L$ links.  Again we will start with $r \, V$ vertices.  Immediately we will connect the vertices into $r$ separate spanning trees.  Each spanning tree will contain $V - 1$ links, and will contain also $2{(L - V + 1)}$ dangling legs which still need to be connected.  However, if the original target theory inhabits a tree, then $2{(L - V + 1)} = 0$ and the perturbation theory stops here; all higher order diagrams are disconnected spanning trees and are deleted by the logarithm in equation \ref{EfetovTheory}.  In this case $Q$ is exactly equal to the partition function of the target theory, precisely as predicted by Efetov, Parisi, and Slanina.

However, if the target theory does not inhabit a tree, there will be an additional phase of connecting the remaining dangling legs, and thus tying the spanning trees together.   This results in many other Feynman diagrams, each corresponding to an alternative geometry, a different way that the $r \, V$ vertices could be connected together.   The logarithm in equation \ref{EfetovTheory} removes disconnected geometries. We now turn to an analysis of the resulting sum.

\subsection{The dominant geometries}
Which geometries dominate the sum?  This question is answered by summing their symmetry factors.   Section \ref{ConnectedWeight} argues that the sum of the symmetry factors of all fully connected geometries of order $r$ is approximately ${(r!)}^{L-V}$.   If the target theory inhabits a $D$-dimensional lattice then ${L - V }= {{(d-1)} V}$,  so this sum is rather large.   For instance, on a $10$x$10$ lattice the second contribution to $Q$,  $Z^T_2$, will be roughly ${(2!)}^{100}$ times as big as $Z^T_1$, the partition function of the target theory.  This large factor is somewhat counterbalanced if the model has a positive free energy density imposing a cost for adding nodes. Nonetheless, because there are geometries with $r$ arbitrarily large, eventually the ${(r!)}^{L-V}$ growth will overwhelm the free energy density. Put another way, the nonextensive growth of the entropy overrides the extensive growth of the free energy.  In summary, large $r$ geometries completely dominate the sum. 

The dominant geometries are locally tree-like, in the sense that the only non-backtracking walks which return to their starting point are infinitely long.  This is obvious from section \ref{SumOverGeometries}'s discussion of how to enumerate geometries, by first constructing $r$ spanning trees and then connecting them in all possible ways.  Each node lies on one of the $r$ different spanning trees.   In order for a non-backtracking walk to come back to its origin, it must necessarily step off of the origin's spanning tree.  Once the walk is on a different tree, each subsequent step will have a probability of order $r^{-1}$ of returning to the origin's spanning tree.   Since large $r$ geometries dominate, the average loop length is infinitely large.

The Bethe approximation is exact on trees, and a good first approximation on geometries which are locally tree-like.  Perhaps this explains, on a physical level, why the Efetov Lagrangian exhibits saddle points related to the Bethe approximation.

\subsubsection{Estimate of the connected symmetry factors} \label{ConnectedWeight}
First of all, the nodes in a geometry are generated by an exponential $e^{\lambda {\mathcal{L}}_1}$;  the $r$ instances of each node in the target geometry are accompanied by a factor of ${(r!)}^{-1}$.   Connecting up all the nodes requires $r L$ links, giving a factor of ${(r!)}^L$.  Therefore the sum of all the order $r$ symmetry factors (of both connected and disconnected geometries) is ${(r!)}^{L-V}$.

I now argue that the proportion of disconnected geometries in this sum is very small. I write $C_r$ for the sum of the symmetry factors of the order $r$ connected geometries, and $D_r$ for the sum of the disconnected symmetry factors.  Thus  ${C_r } = {{(r!)}^{L-V} - D_r}$.  Now the number of disconnected geometries $D_r$ is simply related to the numbers of connected geometries at lower orders:
$D_1 = 0$, $D_2 = {\frac{1}{2} {(C_1)}^2}$, $D_3 = {{C_1  C_2} + {\frac{1}{3!} {(C_1)}^3}}$, $D_4 = {{C_1 C_3} + {\frac{1}{2}{(C_1)}^2 C_2} + {\frac{1}{4!}{(C_1)}^4} + {\frac{1}{2}{(C_2)}^2}}$, et cetera.  I am not aware of any way to write a closed form expression for the total value of $D_r$ except in the special case $L = V$, but its largest single part is the $C_1 C_{r-1}$ term, which is  smaller than ${(r!)}^{L-V}$ by a factor of ${(\frac{1}{r})}^{L-V}$.  All other contributions are much much smaller, of order ${(\frac{1}{r (r-1)})}^{L-V}$ or less.  Therefore I conclude that if $L-V$ is large then the connected weights sum to approximately ${(r!)}^{L-V}$.

\subsubsection{Divergence of Efetov's theory} \label{divergence}
The perturbative expansion around the trivial saddle point diverges. I have already argued that if $L-V$ is large then the total number of order-$r$ geometries is approximately ${{(r!)}^{L-V} }$.  Because the free energy density of each geometry is bounded from above by a geometry-independent value $\textit{f}\,$, the partition function of each geometry is bounded from below by $\exp{(- \beta r  V   \textit{f} \;)}$.  Since the number of geometries ${{(r!)}^{L-V} }$ increases faster than exponentially and each geometry gives a positive contribution, the terms $Z^T_r \approx { {\lambda}^{rV} {{(r!)}^{L-V} } {\exp{(- \beta r  V   \textit{f} \;)}}}$ in the perturbative expansion around the trivial saddle point increase without bound, and the entire series diverges.  The fact that the individual terms in the series are all positive provides strong evidence that not only the perturbative expansion of Efetov's theory diverges but also the theory itself.

This divergence has nothing to do with the volume: it occurs even when the target geometry has only a small number $L$ of links, in which case Efetov's path integral integrates over only a finite number of degrees of freedom.  Instead the divergence is a result of interaction terms in the Efetov lagrangian, much like the way that ${\phi}^4$ theory diverges if the interaction has the wrong sign.  Unlike ${\phi}^4$ theory, the interactions in Efetov's theory are complicated enough that the divergence is not easily visible unless one expands perturbatively in the interaction constant and obtains a sum over geometries.  As we have seen, the perturbative picture of the divergence is that the number of geometries grows factorially, while their free energies grow only exponentially.

Within the framework of the Efetov Lagrangian, the only way to control the divergence is by treating target geometries which have less than two loops or by setting $\lambda = 0$.  Quantum gravity manifests a similar divergence, but it has been regularized by constraining the sum over geometries, either by fixing the topology \cite{Ambjorn95, Loll98} or via Causal Dynamical Triangulations \cite{Ambjorn06}. 

\subsection{Is the geometrical interpretation valid?}
  We have just established that the perturbative expansion around the trivial saddle point diverges, which is standard, since as a rule the perturbative expansions of non-trivial field theories are only asymptotic.  To be more precise about asymptotic theories, their successive perturbative corrections first decrease term by term but eventually start increasing factorially, due to the perturbative expansion's neglect of non-analytic features of the original path integral.  Nonetheless it is well known that the first few terms of an asymptotic expansion can give a very good approximation to the correct result.

There are several reasons to think that Efetov theory's divergence is much more malignant, and to doubt altogether the validity of perturbation theory around the trivial saddle point: 
\begin{itemize}
\item Unlike an asymptotic theory where the coefficients of the perturbative contributions decrease for a while before starting to diverge, this series exhibits exponentially large coefficients already at the second order.  The second order contribution to the trivial saddle point is approximately ${({\lambda} \; {\exp{(- \beta  \textit{f} \;)}})}^V {(2!)}^{L-V} $ bigger than the first order contribution, so that the small parameter $\lambda$ is counterbalanced by a large geometrical factor.  The free energy term $\exp{(- \beta  \textit{f} \;)}$ is freely adjustable by additive contributions to the Lagrangian, and therefore should not be regarded as distinct from the perturbative parameter $\lambda$.
\item The coefficients increase not only factorially, but as a factorial taken to a very large power: ${(r!)}^{L-V}$. 
\item Every order of the expansion contributes with the same sign.
\end{itemize}
The geometrical interpretation of the Efetov Lagrangian - including both its link with the target theory $T$ and its interpretation as a sum over geometries - rests on the validity of this perturbative expansion.  Since this is questionable, the geometrical interpretation must be taken with a grain of salt.  Probably the only way to rigorously justify a connection between the Efetov Lagrangian and any target theory will be to introduce some type of regularization to the Efetov Lagrangian.  

\section{\label{BetheSaddlePoint}THE NON-TRIVIAL SADDLE POINTS}
The previous section discussed the physical meaning of Efetov's theory, which is based on the trivial saddle point.  This section turns to the non-trivial saddle points, which are related to the Bethe approximation.  Efetov, Parisi, and Slanina have already devoted much ink to developing the perturbation theory, in hopes that it will provide systematic corrections to the Bethe approximation.  The focus here is on giving a full account of the saddle point structure, including Goldstone bosons, complete enumeration of the saddle points, computation of the Hessian and its eigenvalues, and the Hessian's nonnormality.

\subsection{\label{GoldstoneBosons} $U(1)$ symmetries and Goldstone bosons}
The Efetov Lagrangian, given in equation \ref{EfetovLagrangian}, possesses many exact $U(1)$ symmetries.  Each of the links can be multiplied by an overall phase: ${\psi}_{vds} \rightarrow {\psi}_{vds} e^{\imath {\phi}_{vd}}$.  There are $L$ links, but only $V$ different linear combinations of the $L$ phases actually appear in the Efetov Lagrangian, one for each term in ${\mathcal{L}}_1$; these linear combinations are the sums $ {\theta}_v \; = \; {\sum_d {{\phi}_{vd} - {\phi}_{v-\hat{d},d} }}$ which obey the constraint $\sum_v {\theta}_v \; = \; 0$.  Therefore the Efetov Lagrangian has $L - V + 1$ exact symmetries.  There is a strict analogy to the following integral with a single exact $U(1)$ symmetry: 
\begin{equation} \label{PhiToTheFourth}
\int {d \psi} {d \overline{\psi}} \; e^{ {-m \psi \overline{\psi}} + {\lambda {(\psi \overline{\psi})}^2} }.
\end{equation}

The exact symmetries create difficulties for the evaluation of non-trivial saddle points.  Goldstone bosons arise from each of them due to spontaneous symmetry breaking.  Goldstone bosons, especially ones caused by exact symmetries, are invitations to integrate exactly, separately from any saddle point approximation.   Indeed, any attempt to include a Goldstone boson in a saddle point approximation will give an infinite result.

\subsection{Exact integration}
The exact $U(1)$ symmetries may be integrated as follows:
\begin{itemize}
 \item Factor out a phase $e^{i {\phi}}$ from each link.  One does this by changing to angular variables ${\psi}_{vds} = {r_{vds} e^{\imath {\phi}_{vds}}}$, adding ${\phi}_{vd, s=0}$ to the other phases (${\phi}_{vd, s \neq 0} \rightarrow {\phi}_{vd, s \neq 0} + {\phi}_{vd, s = 0} $), and factoring out the common phase ${\phi}_{vd, s = 0}$.  A Jacobian term $\sum_{vds} \ln {|{\psi}_{vds}|}$ must be added to the Lagrangian. 
\item Notice that  the Lagrangian depends only on the linear combinations $ {\theta}_v \; = \; {\sum_d {{\phi}_{vd, s=0} - {\phi}_{v-\hat{d},d, s=0} }}$, and therefore integrate out the $L - V + 1$ other linear combinations, which just multiplies Efetov's path integral by a constant Jacobian $J = 1$ and by ${(2 \pi)}^{L - V + 1}$.  
\item Impose the constraint $\sum_v {\theta}_v \: = \: 0$ by inserting a delta function into Efetov's path integral.  Without this constraint the $\theta$ integrations would decouple and the path integral would be exactly zero.
\end{itemize}

The final Lagrangian is
\begin{eqnarray} \label{FinalEfetovLagrangian} \nonumber  \mathcal{L} = &-& {\sum_{v d s_1 s_2}  {\psi}_{v d s_1} {\overline{\psi}}_{v d s_2} {({\gamma}^{-1})}_{s_1 s_2}}  + {\sum_{vds} \ln {|\psi_{vds}|}}
\\ &+& {\lambda {\sum_{vs} {\xi}_{s} e^{\imath {\theta}_v} \;{{\Pi}_d{\psi}_{vds} {\overline{\psi}}_{v-\hat{d}, ds}}}}. 
\end{eqnarray}
The $\psi$'s are in radial coordinates ${\psi}_{vds} = {r_{vds} e^{\imath {\phi}_{vds}}}$, and ${\phi}_{vd, s=0} = 0$.  There is some liberty to convert the ${\psi}_{vd, s \neq 0}$ components back to Cartesian coordinates.  This simplifies the perturbation theory but mutilates the Efetov theory's symmetry with respect to the spin index \footnote{Rather than integrating the $U(1)$ symmetries exactly, one could adopt a different approach: simply remove the Goldstone bosons from the perturbative corrections to the saddle points, by truncating the zero sector of the bare Green's function.  This is equivalent to inserting $L - V + 1$ delta functions into Efetov's path integral, given in equation \ref{EfetovTheory}, and is exactly analogous to inserting a delta function $\delta{(Im \; \psi)}$ into equation \ref{PhiToTheFourth}, transforming it to
$\int {d \psi} \; e^{ {-m {\psi}^2 } + {\lambda {\psi}^4} }$.
Clearly this is an effective way of removing exact symmetries, although it destroys any rigorous connection with the physical content of Efetov's theory, discussed in section \ref{SumOverGeometries}. 
In this approach, the extra ${\hat{\psi}}^{-1}$ term is not added to the saddle point equation. Because Efetov, Parisi, and Slanina do not mention this term, it would appear that they truncated the Goldstone bosons.  However, this is not explicitly mentioned, and Parisi and Slanina instead write of a cancellation of divergences at the one-loop level.  Their neglect of the Hessian's nonnormality, which is discussed in section \ref{NonNormal}, confuses the question still more. 
}.

The presence in the action of $e^{\imath {\theta}_v}$, a complex transcendental function, raises some concern about applying the saddle point approximation to the $\theta$ integrations. Unfortunately exact integration of the $\theta$ variables would add to the Lagrangian a transcendental term of the form ${\sum_{v} \ln f({ \lambda \sum_{s} {\xi}_{s}  \;{{\Pi}_d{\psi}_{vds} {\overline{\psi}}_{v-\hat{d}, ds}}})}$, prohibiting further progress.

\subsection{\label{SaddlePoint}The saddle points}
As is usual when doing mean field theory, we look for a spatially uniform ground state; one that depends on neither the site index $v$ nor the direction index $d$.   It is helpful to first consider the saddle point equation obtained from the final Lagrangian given in equation \ref{FinalEfetovLagrangian} but ignoring the logarithm.  We take the first derivative of the Lagrangian and obtain
\begin{equation} \label{SaddlePointEquations1}
{{\sum_{s_2}   {\hat{\psi}}_{s_2}^{*} {({\gamma}^{-1})}_{s_1 s_2}} = {\lambda { {\xi}_{s_1}  \;{ {\hat{\psi}}_{s_1}^{D-1} {\hat{\psi}}_{s_1}^{* \, D}}}}}.
\end{equation}
The saddle point solutions scale with $\lambda$ as ${\lambda}^{-1/{(2D - 2)}}$; if we remove  $\lambda$ and constrain $\hat{\psi}$ to be real then the saddle point equation becomes the Bethe-Peierls approximation: 
\begin{equation} \label{BetheApproximation}
{   {\hat{\psi}}_{s_2} } = {{\sum_{s_1} {\gamma}_{s_2 s_1} { {\xi}_{s_1}  \;{ {\hat{\psi}}_{s_1}^{2D-1} }}}}.
\end{equation}

I have set $\theta$ to zero because this is the only spatially uniform value which will satisfy the global constraint.  On the other hand, the Lagrangian's first derivative with respect to ${\theta}_v$ is  $\imath \lambda \sum_s {\xi}_s  {{\hat{\psi}}_s}^{2D}$; not only non-zero but also imaginary.  The real part of the first derivative vanishes at the non-trivial saddle points, but the imaginary part does not.  Perhaps this is OK, since the imaginary part contributes only a phase.  As a consequence of this difficulty, $\theta$ will have a vacuum expectation value.  However,  the global constraint on $\theta$ removes the zero-momentum $\theta$ mode from the Hessian, so the vacuum expectation value has no influence on the perturbation theory.

Equation \ref{SaddlePointEquations1} is almost the same as the saddle point equation which is obtained from the original Efetov Lagrangian in equation \ref{EfetovLagrangian}.  The difference is that the saddle point equation before $U(1)$ integration has a $U(1)$ symmetry which must be spontaneously broken by hand, while after integration the phase of ${\hat{\psi}}_{s = 0}$ is constrained to be zero and the $U(1)$ symmetry is no longer present.  Other than this procedural detail, the two equations, and their solutions, are exactly the same.

Equation \ref{SaddlePointEquations1} contains $2S$ equalities which must be satisfied, where $S$ is the number of values that the spin can take.  Since the imaginary part of ${\hat{\psi}}_{s = 0}$ is zero, there are only $2S - 1$ unknowns; one should not expect complex solutions.  On the other hand, real solutions are possible, since in this case there are only $S$ equalities and $S$ unknowns. In particular, each of ${\hat{\psi}}$'s components, with the exception of ${\hat{\psi}}_{s = 0}$, may have either a negative or positive value.  Some solutions of the Bethe approximation actually do have negative values, even though this conflicts with the physical reasoning leading up to the Bethe approximation, in which ${\hat{\psi}}_{s}$ is understood as the value of a partition function.

When the target theory is the Ising model, the Bethe approximation has more than one  solution.  I discuss only the ferromagnetic regime where $\beta J > 0$. Two non-zero real solutions are present at all temperatures, one of which gives a negative value to ${\hat{\psi}}_{s = 1}$.  Below a critical temperature two more solutions appear.  As a general result for all target theories on cubic lattices, these Bethe saddle points are not local maxima of the Lagrangian.  This may be verified by examining the Lagrangian's Hessian (see section \ref{TheHessian}) at the Bethe saddle points and noting that it has a positive eigenvalue corresponding to multiplications of the saddle point solution $\hat{\psi}$ by a real constant.  As a consequence, the states corresponding to the Bethe saddle points are not stable; they are not true ground states, and it is impossible to define a valid perturbation theory around them.  Efetov's original program for calculating corrections to the Bethe approximation depended crucially on doing perturbation theory around the Bethe saddle points; the non-existence of such a theory means that the Efetov Lagrangian is useless for Efetov's original goal, though perhaps quite useful for showing a way forward via connections with quantum gravity.

 \subsubsection{Additional saddle points resulting from the logarithm}
If one does not ignore the logarithm in the final Lagrangian shown in equation \ref{FinalEfetovLagrangian}, the resulting saddle point equation  is not quite the Bethe-Peierls approximation any more: 
\begin{equation} \label{BetheApproximation1}
{   {\hat{\psi}}_{s_2} } = {{\sum_{s_1} {\gamma}_{s_2 s_1} { {(\lambda {\xi}_{s_1}  \; {\hat{\psi}}_{s	_1}^{2D-1}  + {\frac{1}{2}  {\hat{\psi}}_{s_1}^{-1} })} }} }.
\end{equation}

When ${\lambda}^{1 / (D-1)}$ is very small, equation \ref{BetheApproximation1} exhibits all the Bethe solutions from equation \ref{BetheApproximation}, plus four more in the particular case of the Ising model.  As ${\lambda}^{1 / (D-1)}$ is increased, the saddle points change continuously and no new ones appear.  The stability of each saddle point remains the same as well, except at large $\lambda$.  Finally at ${\lambda} \approx 0.1$ all of the solutions disappear. The Bethe saddle points all scale as ${\lambda}^{-1/{(2D - 2)}}$, so at these saddle points the last term in equation \ref{BetheApproximation1} is of order ${\lambda}^{1/{(D - 1)}}$ and can be neglected, and the corresponding term in the Hessian is also very small.  Therefore all the Bethe saddle points are unstable. 

We turn to the other four saddle points, in the limit of small ${\lambda}^{1 / (D-1)}$. The extra saddle points all disrupt the balance between the two terms in the original Bethe approximation, so that either ${   {\hat{\psi}}_{s} }$ or $\lambda {\xi}_{s}  \; {\hat{\psi}}_{s}^{2D-1}$ is negligible.  As a result these saddle points have no relation to the Bethe approximation.  Two of them have ${\hat{\psi}}_{s = 0} = \pm {\hat{\psi}}_{s = 1} = \sqrt{(e^{\beta J} \pm e^{- \beta J})/2}$.  These zero-magnetization solutions are notable both because they don't depend on the magnetic field at all and because the $+$ solution is stable.  The $-$ solution is not stable.  There are also two ferromagnetic saddle points where one component scales as ${\lambda}^{-1 / (2D - 2)}$ and the other scales as ${\lambda}^{1 / (2D - 2)}$.  These are both unstable.

One intriguing aspect of saddle points is their dependence on coordinate transformations.  The extra term in equation \ref{BetheApproximation1} is entirely due to a transformation from Cartesian coordinates to angular coordinates.  The same term could be removed again by transforming from $r$ to $w= r^2$.  If one wanted to preserve the Bethe approximation while playing such games, one would need to introduce a term which would not disturb the ${\lambda}^{-1 / (D-1)}$ scaling of the solutions.  Therefore the coordinate transformation would have to depend on $\lambda$.

\subsection{\label{TheHessian}The hessian}

Evaluation of a saddle point and its corrections starts with computing the Hessian (second derivative) of the Lagrangian at that saddle point.  The inverse of the Hessian will be the bare Green's function used in perturbative corrections to the saddle point, while a Gaussian integral incorporating the Hessian will determine the saddle point's weight.  

Because we have removed the $U(1)$ symmetries, we begin using the following degrees of freedom explicitly: $\vec{\psi} \equiv ({ \vec{{r}}, \vec{{\phi}}, \vec{\theta} )}$.  The following rescaling simplifies the Lagrangian's form around the saddle point: ${\psi}_s = {{\hat{\psi}}_s  {(1 +  {r_s / {\rho}_s} )} \exp{(\imath  {\phi}_s / {\rho}_s)}}$, where ${\rho}_s \equiv {  {\xi}^{1/2}_s {\hat{\psi}}^{D}_s}$.  We also rescale $\theta \rightarrow { \theta  / {|\vec{\rho}|}}$.  In these rescaled variables, the Lagrangian is
\begin{eqnarray} \label{MeanEfetovLagrangian} 
\nonumber   &-&  \sum_{v d s_1 s_2}  {\hat{\psi}}_{s_1}  {\hat{\psi}}_{s_2}   {({\gamma}^{-1})}_{s_1 s_2}    {(1 +  {{r}_{v d s_1} / {\rho}_{s_1}})} \; {(1 +  {{r}_{v d s_2} / {\rho}_{s_2}})} 
\\ \nonumber &\times& \;\;\;\;\;\;\;\; {\exp{( {\imath {\phi}_{v d s_1} / {\rho}_{s_1}} - {\imath {\phi}_{v d s_2} / {\rho}_{s_2}} )}} 
\\ \nonumber  & + & \lambda \sum_{vs}  {\rho}_{s}^{2} \; e^{\imath  {\theta}_v / {|\vec{\rho}|}} \; \; {\Pi}_d \;  {(1 +  {{r}_{v d s} / {\rho}_{s}})} \; {(1 +  {{r}_{v-\hat{d}, d s} / {\rho}_{s}})} 
\\ \nonumber &\times& \;\;\;\;\;\;\;\; {\exp{( {\imath {\phi}_{v d s} / {\rho}_{s}} - {\imath {\phi}_{v-\hat{d}, d s} / {\rho}_{s}} )}}  
\\ &+ & \sum_{vds}{(  {\ln {\hat{\psi}}_{s}} + {\ln  {|1 + {r_{vds} / {\rho_{s}} } |}} )}.
\end{eqnarray}
The Hessian is found by extracting all the second order terms from the Lagrangian, and is
\begin{eqnarray}\label{RealImaginaryHessian} 
\nonumber H^{rr} &=& {-    2A {\delta}_{v_1 v_2} {\delta}_{d_1 d_2} } + {\lambda {\delta}_{s_1 s_2} {\delta}_{v_1 v_2}  {(1 - {\delta}_{d_1 d_2})} }
\\ \nonumber &-& {{ {\rho}_{s_1}^{-2}} \;  {\delta}_{s_1 s_2}  {\delta}_{v_1 v_2} {\delta}_{d_1 d_2}}
+ { \lambda {\delta}_{s_1 s_2}  {(S^T + S + S_2) }}, 
\\  \nonumber H^{\phi \phi} &=& {-  2  P A {(1 - {\delta}_{s_1 s_2})} P {\delta}_{v_1 v_2} {\delta}_{d_1 d_2}} 
\\ \nonumber &-& {{2(\lambda - a_{s_1})} P {\delta}_{s_1 s_2} {\delta}_{v_1 v_2} {\delta}_{d_1 d_2}}
\\ \nonumber &-& { \lambda {P} {\delta}_{v_1 v_2}  {(1 - {\delta}_{d_1 d_2})} }  +  { \lambda {P}  {(S^T + S - S_2)}},  
\\ \nonumber 
H^{\theta \phi } &=& {(H^{\phi \theta})}^T = {- \lambda {P {\hat{\rho}}_{s_2} } {({\delta}_{v_1 v_2} - {S } )}},
\\ \nonumber
H^{\theta \theta} &=& {- \lambda {\delta}_{v_1 v_2}},
\\ \nonumber
H^{r \phi} &=& {(H^{\phi r})}^T
\\ \nonumber &=& {\imath \lambda {P} {\delta}_{v_1 v_2}  {(1 - {\delta}_{d_1 d_2})} } 
+  {{ \imath \lambda {P}   {(S^T - S - S_2)}}},
\\ 
H^{\theta r} &=& {(H^{r \theta})}^T = {\imath \lambda {{\hat{\rho}}_s} {( {\delta}_{v_1 v_2} + {  S } )}}. 
\end{eqnarray}
In the above equations  ${\delta}_{s_1 s_2}$ is the Kronecker delta function,  $S$ is the stepping matrix $S_{v_1 d_1, v_2 d_2} = { {\delta}_{v_1, v_2+\hat{d_2}}}  $,  $S_2$ is the two-step matrix $S_{2, v_1 d_1, v_2 d_2} = {  {(1 - {\delta}_{d_1 d_2})} {\delta}_{v_1+\hat{d_1}, v_2+\hat{d_2}}}$,  $A$ is $A_{s_1 s_2} = { {\xi}_{s_1}^{-1/2} {\xi}_{s_2}^{-1/2} {\hat{\psi}}_{s_1}^{1-D} {\hat{\psi}}_{s_2}^{1-D} {({\gamma}^{-1})}_{s_1 s_2}}$, $a$ is $a_{s_1} = { {\xi}_{s_1}^{-1}  {\hat{\psi}}_{s_1}^{1-2D} \sum_{s_2 \neq s_1} {\hat{\psi}}_{s_2} {({\gamma}^{-1})}_{s_1 s_2}}$, and $P$ is the projection operator $P_{s_1 s_2} = {{\delta}_{s_1 s_2} {(1 - {\delta}_{0 \; s_1})}} $.  One must keep in mind the global constraint on $\theta$, and also that $A$, $a$, and $\rho$ all scale with various powers of $\lambda$.

In the zero-momentum sector the Hessian is considerably simplified:
\begin{eqnarray}\label{RealImaginaryHessian0} 
\nonumber H^{rr} &=& {-    2A  {\delta}_{d_1 d_2} } + {2 \lambda {\delta}_{s_1 s_2}   {(2 - {\delta}_{d_1 d_2})} }
\\ \nonumber &-& {{ {\rho}_{s_1}^{-2}} \;  {\delta}_{s_1 s_2}   {\delta}_{d_1 d_2}}, 
\\  \nonumber H^{\phi \phi} &=& {-  2  P A {(1 - {\delta}_{s_1 s_2})} P  {\delta}_{d_1 d_2}} 
\\ &+& {{2  a_{s_1}} P {\delta}_{s_1 s_2} {\delta}_{d_1 d_2}} .    
\end{eqnarray}

\subsection{\label{NonNormal}The hessian is nonnormal and pseudo-hermitian}
The degrees of freedom break into two classifications: $\vec{{\phi}}$ and $\vec{\theta}$, which are always accompanied by a factor of $\imath$, and $\vec{r}$ which occurs alone.  Broken into these sectors, the Hessian has the form
\begin{equation} \label{ExposeNonnormality1}
H = \left [
		\begin{array}{cc} %
			 \alpha  &  \chi  \\
			{\chi}^T  &  \beta 
		\end{array} \right ].
\end{equation}
  where $\alpha$ and $\beta$ are real and hermitian and $\chi$ is imaginary but not hermitian.  Therefore, $H$'s determinant is always real, and $H$ itself is neither hermitian nor anti-hermitian.  

Any matrix in this form is also nonnormal unless ${\alpha \chi} = {\chi \beta}$, which translates to ${{H^{rr} H^{r\phi}} + {H^{r \theta} H^{\theta \phi}}} = {H^{r \phi} H^{\phi\phi}}$ for the case at hand \footnote{The mapping from the Lagrangian to the Hessian is not unique: it is possible to add any antisymmetric matrix to $\alpha$ and $\beta$, and any matrix to $\chi$ as long as it is compensated by a matching subtraction from ${\chi}^T$.  This seems to offer hope of finding an alternate Hessian which is normal.  The system of equations for determining such a Hessian turns out be quadratic in the unknown matrices, with many linear constraints.  If any solution exists, deriving it is not likely to be trivial. }.  This equality is not verified, so in fact the  Hessian is nonnormal at the non-trivial saddle points \footnote{This nonnormality was overlooked in the calculations by Efetov, Parisi, and Slanina, and their subsequent results about the Green's function and about observables beyond the Bethe approximation may be incorrect. In particular, the expansion of the bare Green's function in terms of its eigenvalues and eigenvectors must make an explicit distinction between right and left eigenvectors. }.  

The Efetov Lagrangian's nonnormality results from treating the underlying geometry as a directed rather than undirected graph.  Any theory without a simple symmetry between $\psi$ and $\overline{\psi}$ will have a nonnormal Hessian. Even a free field theory $\mathcal{L} = {-\psi D \overline{\psi}}$ is nonnormal if $[D, D^T] \neq 0$.  

Within the class of nonnormal matrices, the Hessian is specifically a pseudo-hermitian matrix \footnote{Pseudo-hermitian matrices have recently been of much interest in the study of $PT$ symmetric quantum mechanics \cite{Bender98, Mostafazadeh02}, which is an interesting way of including dissipation effects into the standard structure of quantum mechanics.}, which means that the hermitian conjugate of H has a special relation to H itself: $H = C H^{\dagger} C^{-1}$.  In our case $C$ is the  conjugation matrix $ \left [
		\begin{array}{cc} %
			 1  & 0  \\
			0  &  -1
		\end{array} \right ] $.  As a consequence of the pseudo-hermitian property, if the set of eigenvectors of $H$ is complete then all the eigenvalues of H either are real or occur in complex conjugate pairs, and $C$ maps the right eigenvectors to the left eigenvectors \cite{Mostafazadeh02}.  

If the target theory possesses translational invariance, then the Efetov Lagrangian shares the same symmetry.  In this case the Hessian commutes with the translation matrix, and decouples into sectors corresponding to each value of the momentum.   Because the Hessian's nonnormality is a symptom of spatial structure, the Hessian's zero-momentum sector is normal, as seen in equation \ref{RealImaginaryHessian0}.  

\subsubsection{\label{NonNormalSaddlePoints} The nonnormal saddle point integral}
When using the saddle point approximation, the weight of each saddle point is determined by a product of several quantities, among which is the  Gaussian integral
\begin{equation}\label{GaussianIntegral1}  {   {\int {{d \vec{\psi}} } \;  {\exp{( { \frac{1}{2} \vec{\psi} H \vec{\psi} } )}}}}. \end{equation}
Switching to a momentum basis, one obtains a product of Gaussian integrals which have exactly the same form as integral \ref{GaussianIntegral1}.  The zero-momentum integral is normal, but the other integrals are nonnormal.

I am not aware of any published way to do this integral exactly when, as in our case, $H$ is nonnormal.  The standard results for Gaussian integrals cover only cases where either $\psi$ is a Grassman variable or else $\psi$ is a scalar variable and $H$ is a normal matrix.  In the second case a suitable change of variables will factorize integral \ref{GaussianIntegral1} into a multiple of many one-dimensional Gaussian integrals.  The integral is then equal to ${(2 \pi)}^{B/2} \; {\exp{(-\frac{1}{2} {Tr}\,({\ln -H}))} } \; = \; {(2 \pi)}^{B/2} {(\det{(-H)})}^{-1/2}$ if $H$ is negative definite, and diverges otherwise.  In contrast, because our Hessian is nonnormal, its eigenvectors are not orthogonal, and there is no change of variables that will cause integral \ref{GaussianIntegral1} to factorize. 

Theorem: If $H$ is symmetric and negative definite and its eigenvectors are complete and contained in the matrix $W$, then
\begin{equation}\label{NonPerturbativeNonNormalIntegral}
   {\int {{d \vec{\psi}} } \;  {\exp{( { \frac{1}{2} \vec{\psi} H \vec{\psi} } )}}} = {{(-1)}^{\sigma} e^{\imath \omega}  {\frac{{(2 \pi)}^{B/2}}{ {det{(-H)}}^{1/2}}}},
\end{equation}
where $w_i$ are the eigenvalues of $W$,  $\sigma = \frac{1}{4} \sum_i {(sign{(w_i)} - 1)}^2$, and $\omega = \sum_i {phase{(w_i)}}$.

Proof:
$H$  has the decomposition $H = {W \lambda W^{-1}}$, where $\lambda$ is the diagonal matrix composed of $H$'s eigenvalues.  We change coordinates $\psi \rightarrow W \psi$, obtaining ${(-1)}^{\sigma} {\det{(W)}} {\int {{d \vec{\psi}} } \;  {\exp{( { \frac{1}{2} \vec{\psi} W^{\dagger} W \lambda \vec{\psi} } )}}}$.  Now imagine doing the individual integrals one by one, completing the square each time.  This process of completing the squares is mathematically equivalent to computing the $LU$ decomposition of $W^{\dagger}W\lambda$.  $LU$ decomposition means factoring a matrix into two matrices $L$ and $U$, where $L$ is zero above the diagonal, $U$ is zero below the diagonal, and the diagonal elements of $L$ are equal to one \cite{Golub83}.  If the $LU$ decomposition exists then it is unique.  After completing the squares, the $i$-th integration converges if $Re{(U_{ii})} < 0$.  The result of all the integrations is $ {(-1)}^{\sigma} {\det{(W)}} \; {(2 \pi)}^{B/2} \;  {(\prod_i -U_{ii})}^{-1/2}$.  The  product of $-U$'s diagonal elements is just the determinant of $-W^{\dagger}W \lambda$, giving formula \ref{NonPerturbativeNonNormalIntegral}.

The last step is to establish that the $LU$ decomposition exists and that $Re{(U_{ii})} < 0$ by constructing $L$ and $U$ more explicitly.  $W^{\dagger} W$ is positive definite and therefore has a Cholesky decomposition $W^{\dagger}W = G^{\dagger}G$, where $G$ is zero below the diagonal and its diagonal elements are real and positive \cite{Golub83}.  Defining the matrix $g_{ij} = {{\delta}_{ij} G_{ii}^{-1}}$, we obtain $W^{\dagger}W \lambda = G^{\dagger} g \, g^{-1} G \lambda$.  Therefore the $L U$ decomposition exists: $L = G^{\dagger} g$ and $U = g^{-1} G \lambda$.  The diagonal elements of $U$ are $G_{ii}^{2} {\lambda}_i$, so the integral exists if $H$ is negative definite. Q.E.D.

Lemma: If $H$ has the form given in equation \ref{ExposeNonnormality1}, is negative definite, and its eigenvectors are complete, then 
\begin{equation}\label{NonPerturbativeNonNormalIntegral2}
   {\int {{d \vec{\psi}} } \;  {\exp{( { \frac{1}{2} \vec{\psi} H \vec{\psi} } )}}} = {\pm   {\frac{{(2 \pi)}^{B/2}}{ {det{(H)}}^{1/2} }}}.
\end{equation}
In this case the Gaussian integral must be real because $\int {{d \vec{\psi}} }$ is symmetric under the transformation $\vec{\psi} \rightarrow C \vec{\psi}$, which reverses the sign of the imaginary part of $\vec{\psi} H \vec{\psi}$.  The only possible values for the phase are $\pm 1$.  If $H$ is normal then the ambiguity in sign is correctly resolved by taking the absolute value of ${det{(H)}}^{1/2}$, and it seems likely that the same applies to nonnormal $H$.

This theorem does not address the possibility that equation \ref{GaussianIntegral1} might converge even if one or more eigenvalues of $H$ is zero or positive, since it does not exclude the possibility of using a set of coordinates other than $\psi \rightarrow W \psi$. If $H$ is in the form of equation \ref{ExposeNonnormality1} then convergence is assured as long as $\alpha$ and $\beta$ are both negative definite.   Whether violation of this condition ensures divergence remains to be seen.  Of course divergence is assured if the zero-momentum Hessian is not negative definite.

One can pursue a perturbative strategy,  dividing $H$'s real part $R$ which is hermitian from its imaginary part $I$, and treating $I$ as a perturbation:

\begin{eqnarray} \label{PerturbationTheoryForNonNormal1}
\nonumber H = \left [
		\begin{array}{cc} %
			 \alpha  &  \chi  \\
			{\chi}^T  &  \beta 
		\end{array} \right ]
&=& R + I, 
\\&R& = \left [
		\begin{array}{cc} %
			 \alpha  &  0 \\
			0  &  \beta 
		\end{array} \right ]
, \;\; I = \left [
		\begin{array}{cc} %
			 0  &  \chi  \\
			{\chi}^T  &  0 
		\end{array} \right ],
\end{eqnarray}

\begin{eqnarray}\label{operatorsplitting1}  
 \int & {{d \vec{\psi}} }& \;  {\exp{( {\frac{1}{2} \vec{\psi} H \vec{\psi} }  )}} 
\\ \nonumber &=& \left |\, \exp{(\frac{1}{2} {\frac {d}{d\vec{l}}} \cdot I \cdot  {\frac {d}{d\vec{l}}})} \;{\int {{d \vec{\psi}} } \,  {\exp{( {\frac{1}{2} \vec{\psi} R \vec{\psi} } + {\vec{\psi} \cdot   \vec{l} } )}}} \,\right |_{\vec{l} = 0}. 
\end{eqnarray}
Clearly the remaining integral diverges unless $R$ is negative definite.  Summing all orders in perturbation theory, we obtain
\begin{eqnarray}  \label{perturbativenonnormal1}
{(2 \pi)}^{B/2} &\exp&{\textbf{(}-\frac{1}{2} {Tr}\,[{\ln -R}]\;\textbf{)}}  
\\ \nonumber &\times& {\exp{\textbf{(}-\frac{1}{2} {Tr}\,[{-\sum_t \frac{1}{t} {(-I R^{-1})}^t \;}]\;\textbf{)}} } 
\\ \nonumber &\;&\;\;\;\; \approx {(2 \pi)}^{B/2} \; {\exp{\textbf{(}-\frac{1}{2} {Tr}\,[{\ln -H}]\;\textbf{)}} }.
\end{eqnarray} 
This is exactly the same result that one obtains rigorously via diagonalization when $H$ is normal and negative definite, except that all non-perturbative information has been lost.  The perturbation theory suggests that convergence depends on the spectrum of $R$, while we know that if $H$ is normal then convergence depends on $H$, not $R$.   If, as in our case, $H$ is not normal, then perturbation theory gives no indication about whether there might be finite non-perturbative corrections to equation \ref{perturbativenonnormal1}, or even of the conditions for divergence or convergence.  There is reason to be suspicious of equation \ref{perturbativenonnormal1}, since at most of the Efetov theory's saddle points the perturbation is not small.  

\subsection{Spectra}
As I have already mentioned, all but one of the Efetov theory's non-trivial saddle points have  at least one positive eigenvalue in their zero-momentum sector and are therefore guaranteed to diverge.  The single negative definite saddle point has a magnetization of exactly zero, does not respond to an external magnetic field, and is not a solution of the Bethe approximation. This saddle point is negative definite at all momenta, ensuring that integral \ref{GaussianIntegral1} converges.

The spectra of the saddle points which satisfy the Bethe equation all scale linearly with $\lambda$. Therefore the bare Green's function at the Bethe saddle points scales as ${\lambda}^{-1}$.  Since the perturbative vertices are proportional to $\lambda$, any perturbative diagram with more lines than vertices will diverge proportionally to a negative power of $\lambda$.  

There are two saddle points which are constant with respect to $\lambda$, including the stable saddle point.  Their spectra are constant with respect to $\lambda$, as are the bare Green's functions.  Therefore perturbative corrections are controlled by the small parameter $\lambda$, with one power of $\lambda$ for each vertex.

There are also saddle points where one component scales as ${\lambda}^{-1 / (2D - 2)}$ and the other scales as ${\lambda}^{1 / (2D - 2)}$.  The scaling behavior of their spectra is harder to analyze, but there are eigenvalues in the zero-momentum sector which scale with ${\lambda}^{- {D/{(D-1)}}}$, as well as ones that scale with ${\lambda}^{ {D/{(D-1)}}}$.  

\subsection{The free energy density}
The free energy density, neglecting both constants and perturbative corrections, and assuming that the Hessian is negative definite, is
\begin{eqnarray}\label{FreeEnergyDensity}
\nonumber {\beta f} &=& {\ln {|\vec{\rho}|}} - {D \ln{|{\rho}_0|}} + {2D \sum_s {\ln {| {{\rho}_s}|}} }  
\\ &+& {\frac{1}{2V}  {Tr{\ln{(-H)}}}} + {{(D-1)} \lambda {|\vec{\rho}|}^2}.
\end{eqnarray}

If the saddle point is dominated by a specific number of vertices, then the partition function is equal to $\lambda$ to some power and the free energy density $f$ is proportional to $\log \lambda$.  This is the case for the unconditionally stable saddle point, where $\rho$ and $H$ are proportional to $1$, which means that it does not involve any vertices at all, much like the trivial saddle point.  On the other hand, the free energy density at the Bethe saddle points, if the Hessian were negative definite, would be dominated by the last term in equation \ref{FreeEnergyDensity}.  This term scales as ${\lambda}^{-1/(D-1)}$, so these saddle points are superpositions of many different geometries.

\subsection{Reliability of the computations}
Some of the results reported here were obtained non-analytically, through a code which computes the saddle points and their Hessians and eigenvalues.  The temperature, magnetic field, coupling constants, number of dimensions, and lattice size can all be specified.  Many automated tests are built into the code, including computation of the Hessian three different ways.   If the reader wants to reproduce or check these results, the code is available under the GNU public license and may be downloaded from my web site, www.sacksteder.com.

\section{\label{Outlook}CONCLUSIONS AND FURTHER CHALLENGES}
The most exciting result of this paper is that sums over geometries may be the appropriate tool for understanding and improving mean field theory in $D$ dimensions, and in particular for developing systematic corrections to the Bethe approximation.  There is considerable room for designing various sums over geometries in which the lattice geometries are already dominant or else can be smoothly reweighted to be dominant.  If possible, the sums should be finite, be formulated as integrals, and have non-trivial saddle points which are local maxima of the Lagrangian. In this case they would define interesting mean field theories with systematic corrections.  In order to obtain corrections to the Bethe approximation, one would hope to find an ensemble which allows a smooth reweighting, without phase transitions, between lattice and tree-like geometries.  Quantum gravity and string theory may provide exactly the expertise needed to construct the needed ensemble.

The Efetov theory is, to my knowledge, the first example of a bosonic theory with a nonnormal Hessian and Green's function.  Nonnormal field theories hold some promise for studying non-ergodic and non-equilibrium behavior \cite{Trefethen05}, and for the study of dissipation in quantum systems \cite{Bender98, Mostafazadeh02}.  This paper proves for the first time the value of Gaussian integrals with nonnormal kernels, which is a prerequisite for the development of saddle point approximations and perturbation theory.   It also opens a host of questions: 
\begin{itemize}
 \item What are the necessary and sufficient conditions for convergence of a Gaussian integral with a nonnormal kernel?  Can the phases and signs in equations \ref{NonPerturbativeNonNormalIntegral} and \ref{NonPerturbativeNonNormalIntegral2} be simplified? Integrals with kernels in the form given by equation \ref{ExposeNonnormality1} are particularly important because this is the most general form for bosonic Lagrangians that contain no complex constants.  
\item It is well known that nonnormal matrices are not fully characterized by their eigenvalues.  For instance, their spectra can be exceedingly sensitive to small perturbations, and their powers may show a transient behavior which grows much faster than powers of the largest eigenvalue.  Pseudospectra, plots of how the spectrum would vary under small perturbations of the matrix, are recognized as showing much additional information \cite{Trefethen05}.  Do these issues manifest themselves in the structure of perturbative corrections to a nonnormal saddle point?
\item What is the physical meaning of Feynman diagrams in a nonnormal theory where propagators are extremely sensitive to perturbations?  Does self-energy still have a meaning? 
\end{itemize}

This paper also contributes a more detailed analysis of the Efetov theory, culminating in the discovery that it is actually a sum over geometries, and also in strong evidence that it is divergent and in need of regularization, probably similar to regularizations in quantum gravity. Even without regularization, its interpretation as a sum of tree-like geometries argues that further attention should be given to the $\theta$ integrations.  If some way were found to do these integrals non-perturbatively, perhaps the Efetov theory would yield saddle points which are more satisfactory.

\begin{acknowledgments}
Much of this work was done at the Ateneo de Manila University and the Bose Centre in Kolkata.  Thanks are also due to Feng Yuan Ping for his hospitality at the National University of Singapore, and especially to my thesis advisor, Giorgio Parisi, for suggesting this research topic, finding funding for its initial stages, and reading a late draft of this paper and making an important comment.  
\end{acknowledgments}

\bibliography{vincent}

\end{document}